\documentclass[preprint2]{aastex}

 \newcommand{\pref}{\protect\ref}

\shorttitle{\indent \def Persistent Doppler shift oscillations} \shortauthors{Tian et al.}

\begin{document}

\title{Persistent Doppler shift oscillations observed with HINODE/EIS in the solar corona: spectroscopic signatures of Alfv\'enic waves and recurring upflows}

\author{Hui Tian\altaffilmark{1,*}, Scott W. McIntosh\altaffilmark{1}, Tongjiang Wang\altaffilmark{2,3}, Leon Ofman\altaffilmark{2,3}, Bart De Pontieu\altaffilmark{4}, Davina E. Innes\altaffilmark{5}, Hardi Peter\altaffilmark{5}}
\altaffiltext{1}{High Altitude Observatory, National Center for Atmospheric Research, P.O. Box 3000, Boulder, CO 80307; htian@ucar.edu}
\altaffiltext{2}{Catholic University of America, Washington, DC 20064} \altaffiltext{3}{NASA Goddard Space Flight Center, Code 671, Greenbelt,
MD 20771} \altaffiltext{4}{Lockheed Martin Solar and Astrophysics Laboratory, 3251 Hanover St., Org. ADBS, Bldg. 252, Palo Alto, CA  94304}
\altaffiltext{5}{Max Planck Institute for Solar System Research, 37191 Katlenburg-Lindau, Germany}
\altaffiltext{*}{Now at Harvard-Smithsonian Center for Astrophysics, 60 Garden Street, Cambridge, MA 02138}

\begin{abstract}
Using data obtained by the EUV Imaging Spectrometer (EIS) onboard Hinode, we have performed a survey of obvious and persistent (without significant
damping) Doppler shift oscillations in the corona. We have found mainly two types of oscillations from February to April in 2007. One type is
found at loop footpoint regions, with a dominant period around 10 minutes. They are characterized by coherent behavior of all line parameters (line
intensity, Doppler shift, line width and profile asymmetry), apparent blue shift and blueward asymmetry throughout almost the entire duration.
Such oscillations are likely to be signatures of quasi-periodic upflows (small-scale jets, or coronal counterpart of type-II spicules), which may play an important role in the supply of mass and energy to the hot corona. The other type of oscillation is usually associated with the upper part of loops. They are most clearly seen in the Doppler shift of coronal lines with formation temperatures between one and two million degrees. The global wavelets of these oscillations usually peak sharply around a period in the range of 3-6 minutes. No obvious profile asymmetry is found and the variation of the line width is typically very small. The intensity variation is often
less than 2\%. These oscillations are more likely to be signatures of kink/Alfv\'en waves rather than flows. In a few cases there seems to be
a $\pi$/2 phase shift between the intensity and Doppler shift oscillations, which may suggest the presence of slow mode standing waves according to wave theories. However, we demonstrate that such a phase shift could also be produced by loops moving into and out of a spatial pixel as a result of Alfv\'enic oscillations. In this scenario, the intensity oscillations associated with Alfv\'enic waves are caused by loop displacement rather than density change. These coronal waves may be used to investigate properties of the coronal plasma and magnetic field.
\end{abstract}


\section{Introduction}
In the past 15 years there was a rapid growth of literature reporting quasi-periodic oscillations in the upper solar atmosphere \citep[for latest reviews see, e.g.,][]{Banerjee2007,DeMoortel2012b}. These oscillations are often explained as signatures of various modes of MHD waves. For example, \cite{Jess2009} reported an oscillation associated with a large magnetic bright point and interpreted it as torsional Alfv\'en wave. Intensity oscillations of a flare loop observed in radio wavelengths were interpreted as fast sausage mode and applied to constrain the
density contrast between the loop and the background \citep{Nakariakov2003,Aschwanden2004}. Evidence of fast sausage modes was also recently found in magnetic pores \citep{Morton2011}. 

Lateral transverse oscillations are often interpreted as fast-mode kink waves or Alfv\'en waves. The ubiquitous presence of transverse waves has been observed in chromospheric spicules and mottles
\citep{DePontieu2007,Singh2007,Zaqarashvili2007,Kukhianidze2006,Kim2008,He2009b,Okamoto2011,Kuridze2012}, coronal loops
\citep{Aschwanden1999,Nakariakov1999,Nakariakov2001,Verwichte2004,Tomczyk2007,Tomczyk2009,McIntosh2011,Singh2011}, and plume-like structures in different regions of the Sun \citep{McIntosh2011,Tian2011b}. Case studies suggest that some transverse waves are likely associated with impulsive phenomena such as flare activities, filament eruptions, or coronal mass ejections (CME) \citep{Aschwanden1999,Aschwanden2002,Schrijver2002,Wang2004,OShea2007b,Chen2008,Ofman2008,He2009a,Verwichte2010b,Aschwanden2011,Pietarila2011,Liu2011,Liu2012,Shen2012,White2012,White2012b,Wang2012b}. Large-scale transverse waves with clear signatures of damping have also been reported in coronal streamers and found to be driven by CME impingement \citep{Chen2010,Chen2011,Feng2011}. Modeling efforts have also been made to understand the generation and propagation of these transverse waves \citep[e.g.,][]{Matsumoto2010,Murawski2010a,Soler2011,Selwa2010,Selwa2011a,Selwa2011b} and we refer to a recent review of \cite{Ofman2009}. Observations of these Alfv\'enic waves \citep{Goossens2009}, either persistent or impulsively excited, have been widely used for the derivation of coronal Alfv\'en speed and magnetic field. For reviews of the observations of these transverse waves and their applications in coronal seismology, we refer to \cite{Nakariakov2005}, \cite{Erdelyi2008a} and \cite{Zaqarashvili2009}. In spite of the implications of such transverse waves and their role in the MHD seismology of the solar corona, the detection of such waves may also be important for diagnosing the stellar coronae \citep[e.g.,][]{Mitra-Kraev2005,Pandey2009}.

Longitudinal oscillations have also been intensively studied. Using mainly broadband imaging observations, quasi-periodic upward propagating disturbances with a period of three to thirty minutes were frequently observed in polar plumes \citep[e.g.,][]{Ofman1997,Ofman1999,Ofman2000,DeForest1998,KrishnaPrasad2011} and coronal loops
\citep[e.g.,][]{Berghmans1999,DeMoortel2000,DeMoortel2002,Robbrecht2001,Marsh2003,Marsh2009,King2003,DePontieu2003,Ugarte2004a,Ugarte2004b,DePontieu2005,McEwan2006,Tian2008,Srivastava2008,Srivastava2010,Stenborg2011,Yuan2012}. Propagating disturbance with a shorter period (50 s) has been recently reported by \cite{LiuJiajia2012} in a dark cavity region. Longitudinal disturbances have also been investigated through spectroscopic observations
\citep[e.g.,][]{Banerjee2000,OShea2007a,Banerjee2009,Gupta2009,Gupta2010,Wang2012a,KrishnaPrasad2012}, which often reveal a correlation between changes of the line intensity
and Doppler shift \citep{Brynildsen1999,Wikst2000,Wang2009a,Wang2009b,Kitagawa2010,Mariska2010,Gupta2012}. These long-lasting propagating disturbances
often have a speed of 50-200~km~s$^{-1}$ and are usually interpreted as slow-mode magneto-acoustic waves propagating into the corona along the magnetic
field lines. \cite{DePontieu2004} proposed that some of these longitudinal oscillations might be caused by the leakage of p-modes into higher
layers of the solar atmosphere along inclined magnetic field lines. Longitudinal oscillations with clear signatures of
damping have also been detected \citep{Wang2002,Wang2003a,Wang2003b,Wang2005,Mariska2008}. In some cases there is a $\pi$/2 phase shift between
the oscillations of the intensity and Doppler shift \citep{Wang2003a,Wang2003b,Mariska2008}, suggesting the possible presence of standing slow
waves. There are suggestions that these longitudinal oscillations can be used to infer thermal conduction \citep{Ofman2002b}, adiabatic index \citep{VanDoorsselaere2011} and magnetic field \citep{Wang2007} in coronal loops. For reviews of the observations and theories of these long-lasting longitudinal propagating
disturbances and quickly damped longitudinal oscillations, we refer to \cite{DeMoortel2009}, \cite{Taroyan2009}, \cite{Banerjee2011} and
\cite{Wang2011}.

However, \cite{DePontieu2010} and \cite{Tian2011a} argued that quasi-periodic signals are not necessarily waves. They found that in loop footpoint
regions all of the four line parameters (line intensity, Doppler shift, line width and profile asymmetry) show coherent behavior, which can be
easily explained by the scenario of recurring upflows (or quasi-periodically enhanced and weakened upflows). In addition, \cite{Tian2011a} found
a net blue shift and blueward asymmetry of the line profiles at the boundary of an active region (AR) throughout the entire observation
duration, favoring the interpretation of a recurring high-speed (50-150~km~s$^{-1}$) outflow superimposed on the nearly static coronal background. The
presence of these upflows has been confirmed through detailed analysis of emission line profiles
\citep{Hara2008,DePontieu2009,McIntosh2009a,McIntosh2009b,Peter2010,Bryans2010,Martnez-Sykora2011,Dolla2011,Ugarte-Urra2011,Tian2011a,Tian2011c,Tian2012,Doschek2012}. \cite{Tian2012} also reported similar high-speed upflows in CME-induced dimming regions. Based on the similarities of the upward propagating disturbances in plumes and AR boundaries, \cite{McIntosh2010a} and \cite{Tian2011b} suggested that the propagating
disturbances in plumes might also be dominated by flows, although an accurate investigation of the line profile asymmetry in coronal holes is impossible with
current instruments. If this scenario is correct, these high-speed episodic upflows/jets might be an efficient means to provide heated mass into the corona
and solar wind \citep{Sakao2007,DePontieu2009,McIntosh2009b,Hansteen2010,DePontieu2011,Tian2011b}. Recurrent plasma ejections and X-Ray jets have also been identified from recent imaging observations \citep{Morton2012,Zhang2012,Su2012}. Several numerical studies have been performed to investigate the formation of these plasma ejecta \citep[e.g.,][]{Murawski2010b,Murawski2011,Srivastava2011,Martnez-Sykora2011,McLaughlin2012}.

The discovery of these rapid quasi-periodic upflows (or jets, plasma ejections) from spectroscopic observations challenges the universal wave interpretation for coronal
oscillations, specifically for some longitudinal propagating disturbances. There is no doubt that both waves and flows exist in the upper solar
atmosphere. But it is difficult to distinguish between them without analyzing all line parameters including intensity, Doppler shift, line width
and profile asymmetry \citep{DePontieu2010,Tian2011a}, although realistic numerical MHD models may help resolve this ambiguity \citep{Ofman2012}. In this paper, we perform a survey of obvious and persistent (without significant damping)
Doppler shift oscillations in the corona by using the data obtained by the EUV Imaging Spectrometer \citep[EIS,][]{Culhane2007} onboard Hinode.
Through detailed analysis of the temporal evolution of all the line parameters, we conclude that some oscillations are better explained by
recurring upflows and others are more likely to be real waves.

\section{Data selection and reduction}
EIS made a large number of sit-and-stare (fixed slit location) observations from February to April in 2007. We examined the quicklook Dopplergrams (automatically generated level 2 data) of the
Fe~{\sc{xii}}~195.12\AA{} line, which are available on the website of Hinode Science Data Center Europe, for these observations. In
Table~\pref{tab.1} we list the starting time of all observations in which we found obvious (easily identified by eye) and persistent (without significant damping) Doppler
shift oscillations. The locations of oscillations on the slits are also listed in Table~\pref{tab.1}. Typical cadence of these observations is
30~s.

After standard correction and calibration of the EIS data, a running average over 3 pixels along the slit was applied to the spectra to improve
the signal to noise ratio. The spatial offset in the Solar-Y direction between the two CCDs have been corrected using the standard EIS routine 
{\it eis\_ccd\_offset.pro}. This routine measures the offset by co-aligning images from Fe~{\sc{viii}}~185.21\AA{} and Si~{\sc{vii}}~275.35\AA{}. We mainly selected three strong emission lines in each observation for our study: Fe~{\sc{xii}}~195.12\AA{},
Fe~{\sc{xiii}}~202.04\AA{}, and Fe~{\sc{xiv}}~264.78\AA{}. The latter two lines are believed to be clean \citep{Young2007}, although an
unidentified line might be present in the red wing of Fe~{\sc{xiii}}~202.04\AA{} \citep{McIntosh2010b,Tian2012}. The Fe~{\sc{xii}}~195.12\AA{}
line is known to be blended with the Fe~{\sc{xii}}~195.18\AA{} line that is usually a few percent of the 195.12\AA{} intensity
\citep{Young2007,Young2009,Brown2008}. We used this line since it is the strongest EIS coronal line in non-flare conditions. The Fe~{\sc{xii}}~195.12\AA{} line was also involved in density diagnostics and the coalignment between TRACE and EIS. More lines were selected
for a comprehensive analysis of the 2007 March 28 observation.

As a common practice, we applied a single Gaussian fit to each EIS spectrum to derive the line intensity, Doppler shift and line width. Since
there are no cool chromospheric lines in the EIS spectral range, we simply assumed a zero average Doppler shift of each line in each observation.
In some observations there was significant jitter in the solar-y direction. Similar to \cite{Kitagawa2010}, we performed a cross-correlation
between intensities of Fe~{\sc{xii}}~195.12\AA{} at different exposures to evaluate and remove the jitter in the y-direction. It turned out that the jitters were mostly smaller than 2$^{\prime\prime}$. Since the oscillations we found are associated with specific locations on the slit, we can safely conclude that they are not caused by possible jitters in the x-direction. We also calculated the average RB asymmetry in
the velocity range of 60-120 km~s$^{-1}$ for each line profile. The RB asymmetry technique was first introduced by \cite{DePontieu2009} and
later modified by \cite{Tian2011c}. The RB asymmetry here refers to RB$_{P}$, where the spectral position corresponding to the maximum intensity is used as the line centroid \citep[see details in][]{Tian2011c}.

\section{2007 March 28 oscillations: two types of oscillations observed on the same slit}
\begin{figure*}
\centering {\includegraphics[width=0.8\textwidth]{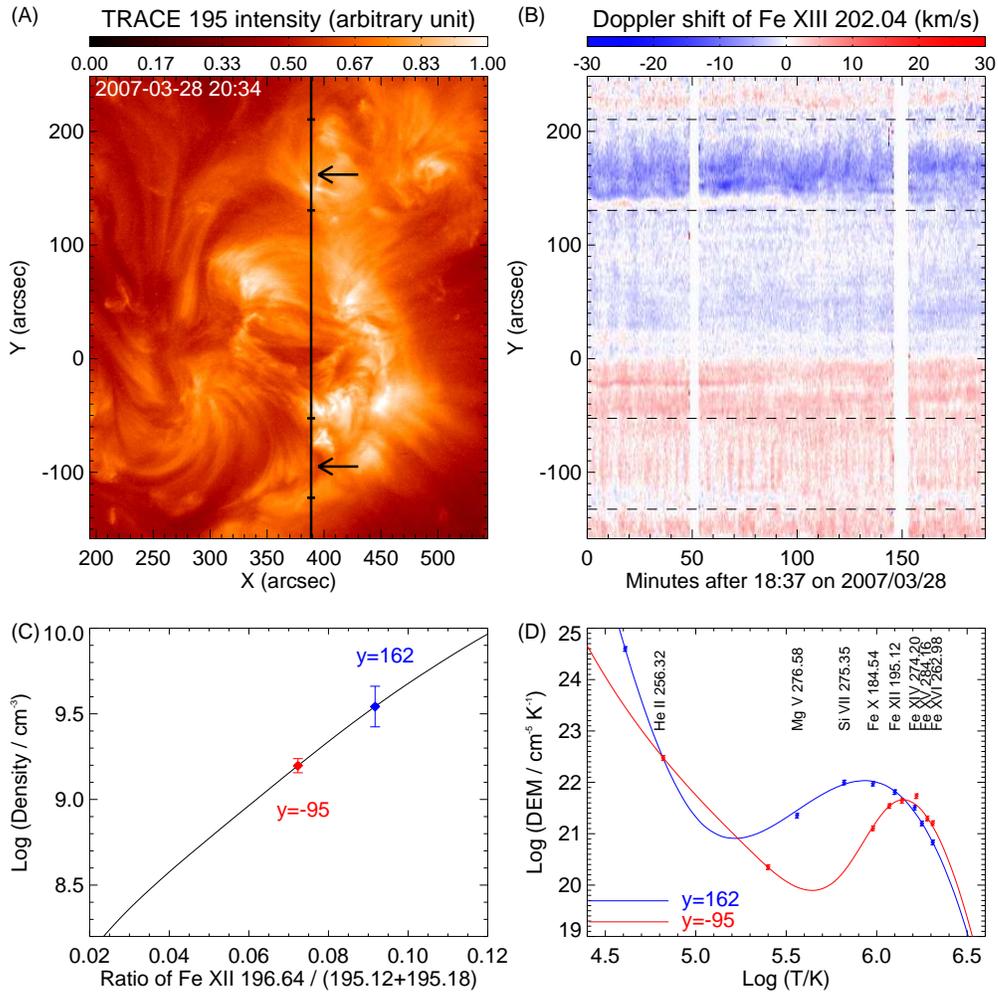}} \caption{ (A) A TRACE 195\AA{} image taken at 20:34 on 2007 March 28. The black
vertical line indicates the location of the EIS slit. The arrows point to the approximate locations of two persistent oscillations. (B) Temporal evolution of the Doppler shift of Fe~{\sc{xiii}}~202.04\AA{} along the slit. The short horizontal bars in (A) and dashed lines in (B) are used to indicate the ranges of illustration in Figs.2-5. (C) Electron densities at the positions of y=162$^{\prime\prime}$ and y=-95$^{\prime\prime}$, respectively. Error bars indicate the standard deviations of the density time series. (D) DEM curves at the positions of y=162$^{\prime\prime}$ and y=-95$^{\prime\prime}$, respectively. }
\label{fig.1}
\end{figure*}

Starting from 18:37 on 2007 March 28, EIS performed sit-and-stare observation in an AR continuously for more than three hours.
Figure~\ref{fig.1}(A) shows an image of this AR in TRACE 195\AA{} passband and the location of the EIS slit as determined by cross-correlating
the EIS Fe~{\sc{xii}}~195.12\AA{} line intensity along the slit and the TRACE intensity at different x-locations. From the temporal evolution of
the Doppler shift of Fe~{\sc{xiii}}~202.04\AA{} shown in Figure~\ref{fig.1}(B), we can clearly see  two locations (indicated by the two arrows) with obvious oscillations throughout the entire duration. One oscillation (around y=162$^{\prime\prime}$) is characterized by some inclined elongated features of enhanced blue shift. This oscillation appears to be associated with the footpoints (or lower parts) of AR loops, as can be seen from Figure~\ref{fig.1}(A). We call
it oscillation 1 in the following discussion. The other oscillation (arround y=-95$^{\prime\prime}$) is dominated by small red shift
and we can clearly see the quasi-periodic appearance of vertical features with enhanced red shift. From Figure~\ref{fig.1}(A) we can see that
this oscillation (oscillation 2) appears to be associated with the tops (or upper parts) of a bundle of AR loops. The slit is found to be almost parallel to the loop plane and cover at least 45$^{\prime\prime}$ of the loop top. Note that although the edge the bright region around y=-70$^{\prime\prime}$ is caught by the slit, this part does not show significant oscillation and it is not analyzed in the following. 

We have performed plasma diagnostics for these two regions revealing clear Doppler shift oscillations. As an example, Figure~\ref{fig.1}(C)\&(D)
show the electron densities and differential emission measure (DEM) curves at the positions of y=162$^{\prime\prime}$ and
y=-95$^{\prime\prime}$. The densities were derived from the intensity ratio of the line pair Fe~{\sc{xii}}~196.64/195.12\AA{}. The theoretical
relationship between the line ratio and density, as extracted from the CHIANTI database \citep{Dere1997,Landi2006}, is shown as the black line.
The electron densities at the two locations are log ({\it N}$_{i}$/cm$^{-3}$)=9.54$\pm$0.12 and 9.20$\pm$0.04, respectively. More lines were used for the derivation of DEM curves. We averaged the profiles of each line acquired at different exposures and assumed an
uncertainly of 10\% for the intensities \citep{Tripathi2008}. By using the routine {\it chianti\_dem.pro} \citep[also used
by][]{Lee2011,Tian2012} available in SolarSoft (SSW) and assuming a constant pressure of 10$^{16}$~cm$^{-3}$~K, we obtained the DEM curves at
the two locations. The peak temperatures at the two locations are around log ({\it T}/K)=5.85 and 6.15, respectively. 

\subsection{Oscillation 1: correlated changes of all line parameters}
\begin{figure*}
\centering {\includegraphics[width=\textwidth]{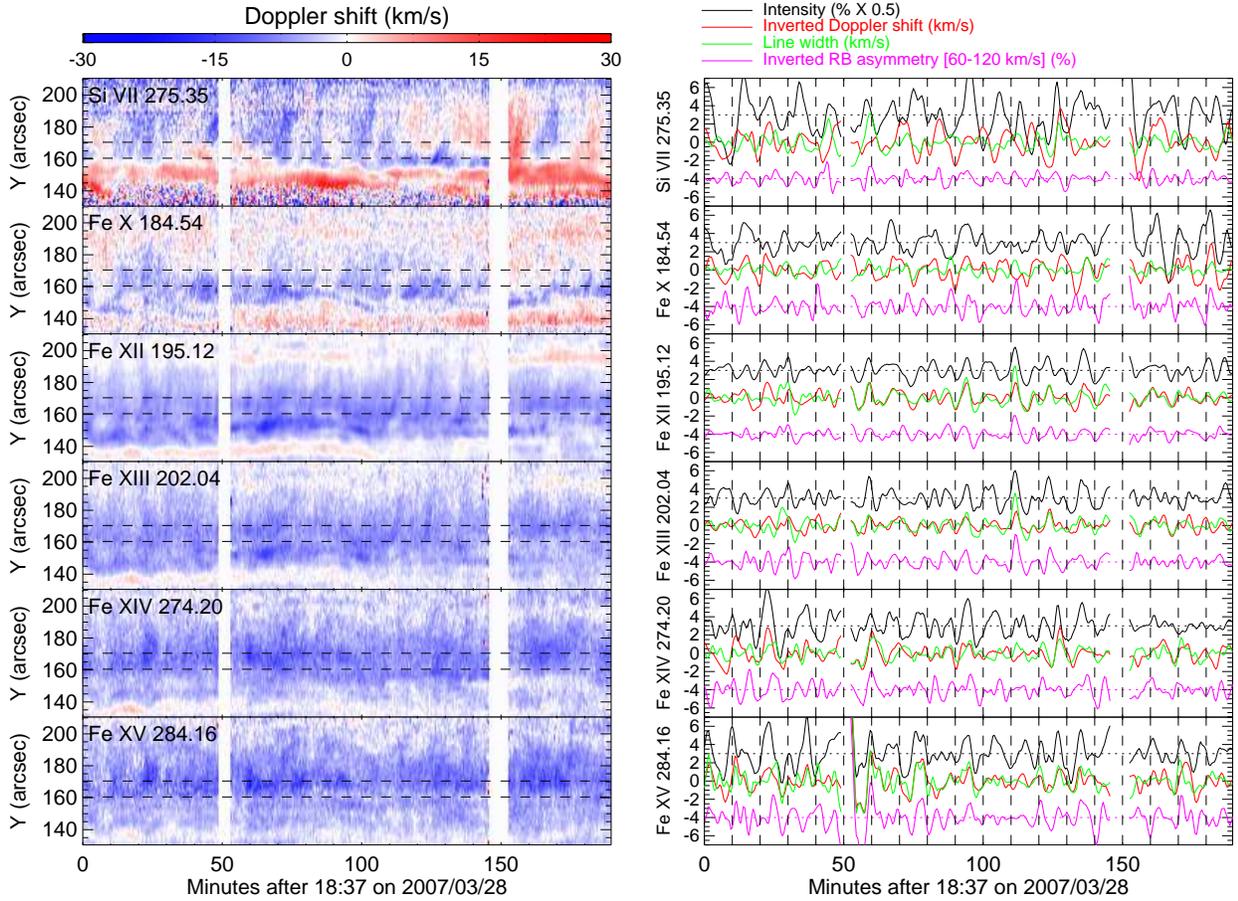}} \caption{ Left: Temporal evolution of the Doppler shift of several emission lines in the
range of y=130$^{\prime\prime}$$\sim$210$^{\prime\prime}$ in Figure~\ref{fig.1}(B). Right: The black, red, green, and violet curves represent
respectively the detrended line intensity, Doppler shift (inverted), line width, and RB asymmetry (inverted) averaged over the region between the two dashed lines shown in the left panels. The intensity has been normalized to the local background and is shown as the percentage divided by 2. The RB asymmetry has
been normalized to the peak intensity of the line profile and is shown in the unit of percentage. The units of Doppler shift and line width are
km~s$^{-1}$. For the purpose of illustration, the intensity and RB asymmetry are offset by 3 and -4 respectively on the y-axis.} \label{fig.2}
\end{figure*}

These oscillations are also found in several other emission lines with different formation temperatures. From the left panels of
Figure~\ref{fig.2} we can see that oscillation 1 is clearly present in all lines with a formation temperature of 0.6-2.0 MK. The inclined
elongated features of enhanced blue shift are most clearly seen in the strongest line Fe~{\sc{xii}}~195.12\AA{} and can also be identified in
other weaker lines. For all coronal lines, we see blue shift throughout almost the entire observation duration. The transition region (TR) line
Si~{\sc{vii}}~275.35\AA{} reveals a mixture of both blue shift and red shift, which might be related to the enhancement of downflows at lower
temperatures \citep{Tian2011c,McIntosh2012}. We averaged all the line parameters (intensity, Doppler shift, line width and RB asymmetry) over
the 11 pixels between the two dashed lines and produced a time series of each line parameter. By subtracting a 10-minute running average from
each time series \citep[e.g.,][]{Wang2009a} we obtained a time series of each detrended line parameter, which is shown in the right panels of Figure~\ref{fig.2}. We found
that the intensity variation generally follows the variation of the line width. It is also clear that the Doppler shift and RB asymmetry show
coherent behaviors. By taking the inverted values of Doppler shift and RB asymmetry (now positive values mean blue shifts or blueward asymmetries), we found that all of the four parameters show coherent variations. Note that the less-than-ideal correlations at some instances are probably partly caused by the poor spectral resolution and photon noise of the EIS instrument \citep{DePontieu2010,Tian2011a}. As demonstrated by \cite{DePontieu2010}, the photon noise has progressively worse effects as one goes to higher moments, i.e., the asymmetry suffers much more from noise than the intensity. The foreground and background emission in the line of sight (LOS) direction might be another reason for the
reduced correlations.

\begin{figure*}
\centering {\includegraphics[width=0.9\textwidth]{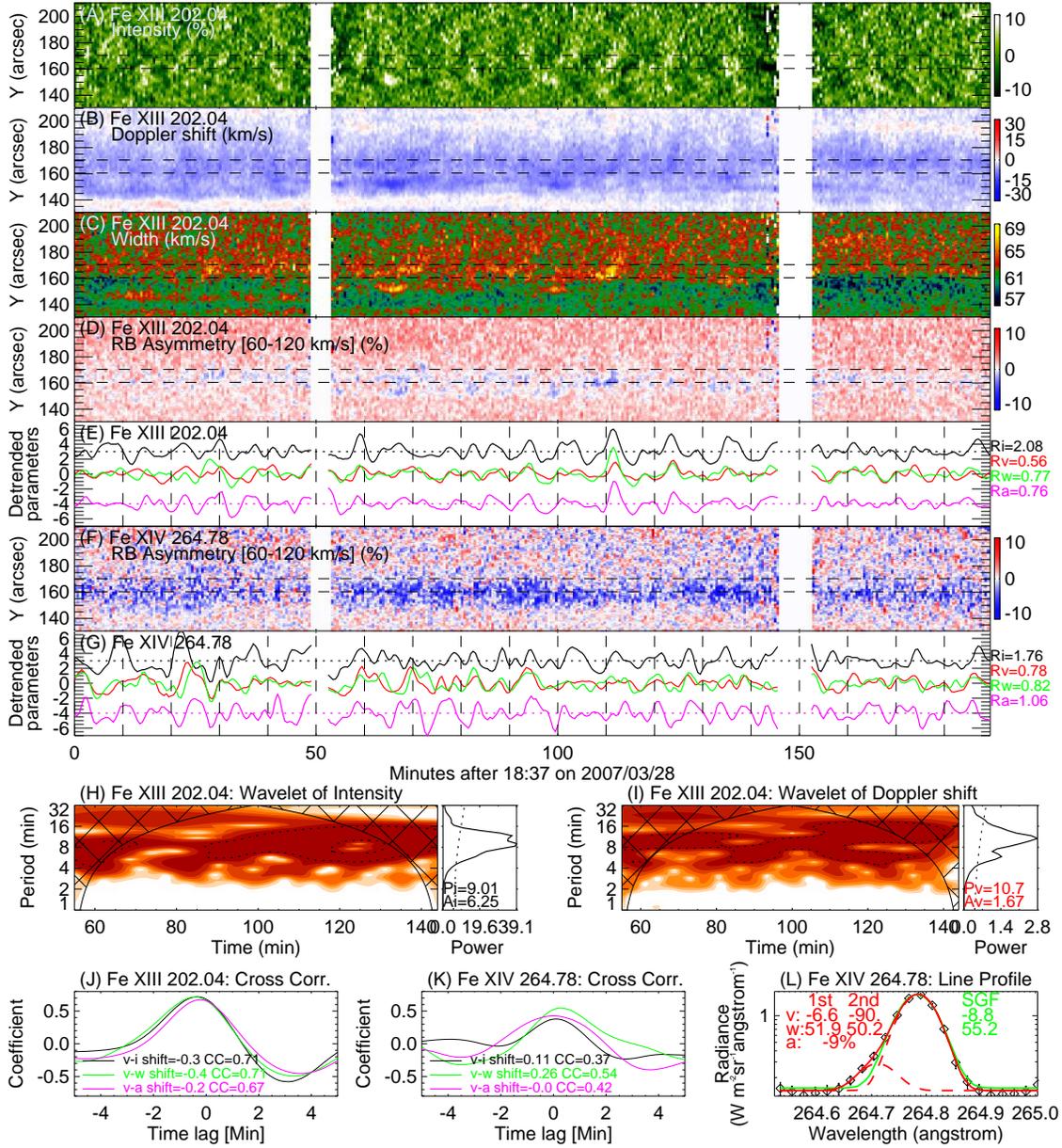}} \caption{ (A)-(D) Temporal evolution of the line parameters (intensity: detrended, others: original) of
Fe~{\sc{xiii}}~202.04\AA{} in the range of y=130$^{\prime\prime}$$\sim$210$^{\prime\prime}$ in Figure~\ref{fig.1}(B). (E) Temporal evolution of
the detrended line intensity (i), Doppler shift (v), line width (w), and RB asymmetry (a) averaged over the region between the two dashed lines
shown in panels (A)-(D). The line styles, units, and illustration methods are the same as any right panel in Figure~\ref{fig.2}. The
root-mean-square values of these parameters are also shown to the right. (F) Similar to (D) but for Fe~{\sc{xiv}}~264.78\AA{}. (G) Similar to (E) 
but for Fe~{\sc{xiv}}~264.78\AA{}. (H)-(I) Wavelet spectra for the detrended intensity and Doppler shift of Fe~{\sc{xiii}}~202.04\AA{}. The periods and
amplitudes derived from the global wavelets are indicated as Pi/Pv and Ai/Av, respectively. The dashed lines indicate a significance level of 99\%. (J)-(K) Cross correlations between the
Doppler shift and intensity(black)/line width(green)/RB asymmetry(violet) for Fe~{\sc{xiii}}~202.04\AA{} and Fe~{\sc{xiv}}~264.78\AA{}. The maximum 
correlation coefficients (CC) and the corresponding time lags (shift) are also marked. (L) An observed Fe~{\sc{xiv}}~264.78\AA{} line profile and 
measurement errors are shown as the diamonds and error bars, respectively. The green line is the single Gaussian fit. The two dashed red lines represent the two Gaussian components and the solid red line is the sum of the two. The velocity (v) and exponential width (w) derived from the single (SGF) and double (1st/2nd for the two
components) Gaussian fits are shown in the panel. Also shown is the intensity ratio of the two components (a).} \label{fig.3}
\end{figure*}

The detailed temporal evolutions of all of the four line parameters of the strong Fe~{\sc{xiii}}~202.04\AA{} line are presented in Figure~\ref{fig.3}(A)-(D). The
loop footpoint region is clearly characterized by quasi-periodic enhancement of the line intensity, blue shift and line width, similar to what \cite{DePontieu2010} and \cite{Tian2011a} found. We can also see weak signatures of blueward 
asymmetry in the region between the two dashed lines. The Fe~{\sc{xiv}}~264.78\AA{} line is clean and from panel (F) we can clearly see obvious blueward asymmetry in the loop footpoint region. A comparison between panels (D) and (F) suggests that the Fe~{\sc{xiii}}~202.04\AA{} line is possibly affected by an unidentified blend (and/or
slight gradient in the background emission) at its red wing \citep{DePontieu2010,McIntosh2010b,Tian2012}. This effect tends to shift the RB asymmetry to the redward and thus reduce the magnitude of blueward asymmetry in the loop footpoint region.  
However, we still used the Fe~{\sc{xiii}}~202.04\AA{} line for our detailed analysis since it is much stronger than Fe~{\sc{xiv}}~264.78\AA{}. The similar behavior of different coronal lines as shown in Figure~\ref{fig.2} suggests that the effect of the possible blend or background issue is not significant in our study. Panel (L) presents a typical line profile of Fe~{\sc{xiv}}~264.78\AA{} (nine profiles averaged) and we can clearly see an enhancement of the blue wing. Such an asymmetric line profile suggests at least two emission components
\citep{Hara2008,DePontieu2009,DePontieu2010,McIntosh2009a,McIntosh2009b,Peter2010,Bryans2010,Martnez-Sykora2011,Dolla2011,Ugarte-Urra2011,Tian2011a,Tian2011c,Doschek2012} and a study of center to limb variation clearly suggests that the enhancement in the blue wing is not caused by blends or noise \citep{Tian2012}.
The correlated changes of different line parameters can be seen from the detrended time series in both panels (E) and (G). Panel (J) and (K) show the cross
correlations between the Doppler shift and intensity/line width/RB asymmetry as a function of time lag. The nearly zero time lag for each of the
parameter pair confirms the coherent behaviors of all line parameters. Similar to \cite{Wang2009a}, we measure the oscillation period as the
value corresponding to the peak global wavelet power and the uncertainty as the half width at half maximum (HWHM). The oscillation periods of the intensity and Doppler shift are 9.01$\pm$2.64 min and 10.7$\pm$2.70 min respectively. The Doppler shift period of 10.7$\pm$2.70 min is consistent with the typical time gap between two inclined stripes in the left panels of Figure~\ref{fig.2}. The large uncertainties of periods suggest that the time gap between two events (i.e., upflows, see below) varies significantly with time. The oscillation amplitude, defined by the square root of the peak global wavelet power \citep{Wang2009a}, is 6.25\% for the intensity and 1.67 km~s$^{-1}$ for the Doppler shift. Note that the data gaps separate the entire time series into three segments and that we selected the longest segment for wavelet analysis. We have also performed the wavelet analysis for the other two shorter segments and the obtained periods are close to that of the central segment. 

\subsection{Oscillation 2: oscillation most clearly seen in Doppler shift}
\begin{figure*}
\centering {\includegraphics[width=\textwidth]{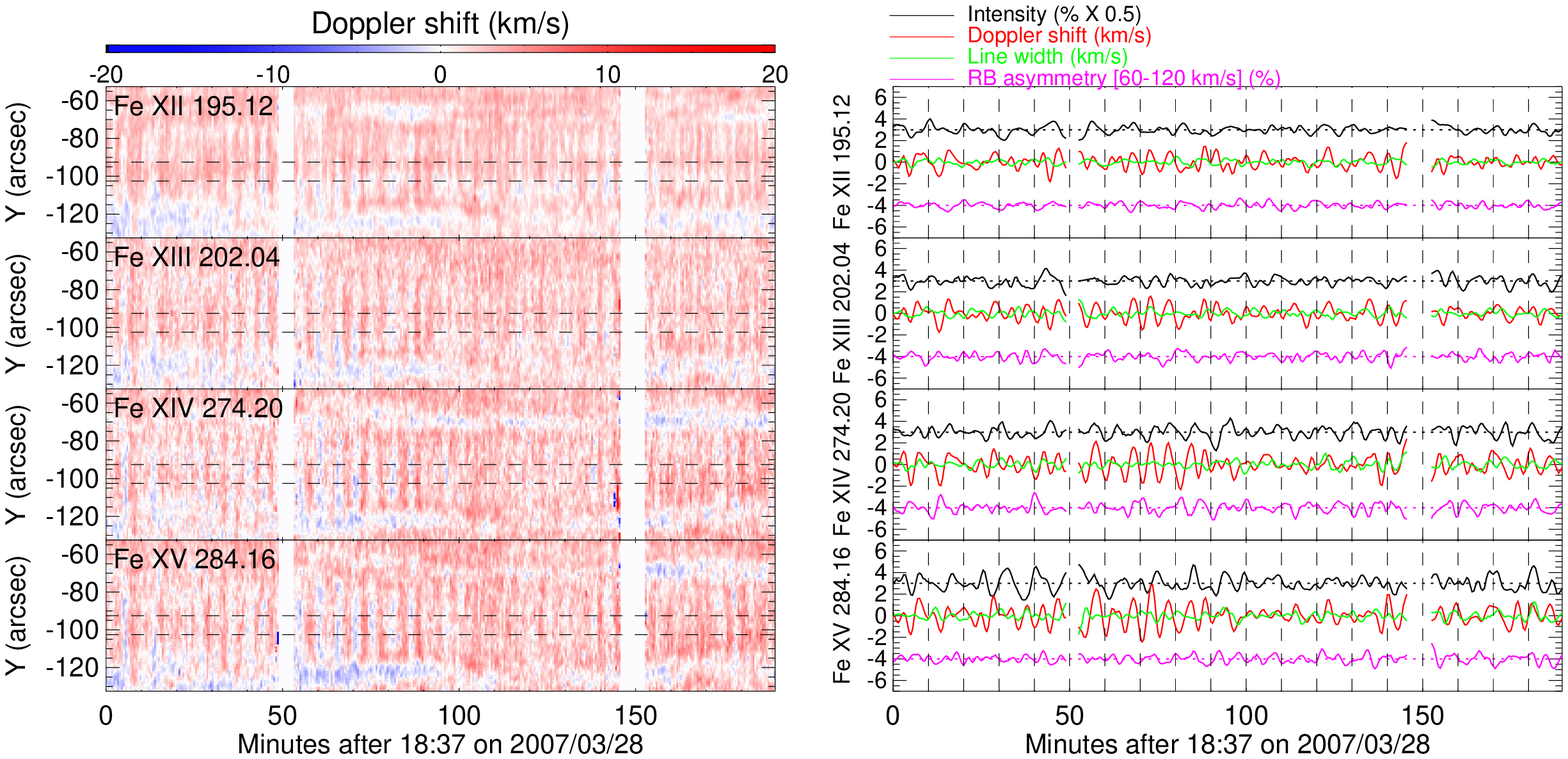}} \caption{ Similar to Figure~\ref{fig.2} but for the range of
y=-133$^{\prime\prime}$$\sim$-53$^{\prime\prime}$ in Figure~\ref{fig.1}(B). The Doppler shift and RB asymmetry values are not inverted here, which is also different from Figure~\ref{fig.2}.} \label{fig.4}
\end{figure*}

Oscillation 2 can be clearly identified in all lines with a formation temperature ranging from 1.3 to 2.0 MK. From the left panels of
Figure~\ref{fig.4} we can see that oscillation 2 is characterized by almost vertical stripes of enhanced red shift. To the best of our
knowledge, such obvious long-lasting oscillations in warm (1.3-2.0 MK) coronal lines have never been found in observations before.
Lower-temperature lines such as Si~{\sc{vii}}~275.35\AA{} and Fe~{\sc{x}}~184.54\AA{} are weak in this loop top region so that we could not
study the detailed temporal evolution of their Doppler shift. Similar to Figure~\ref{fig.2}, we obtained time series of the detrended line
parameters for each line in the region between the two dashed lines and present them in the right panels of Figure~\ref{fig.4}. It is clear that
the oscillation is most prominent in Doppler shift. The variations of the intensity, line width and RB asymmetry are all typically smaller than
those of oscillation 1.

\begin{figure*}
\centering {\includegraphics[width=0.9\textwidth]{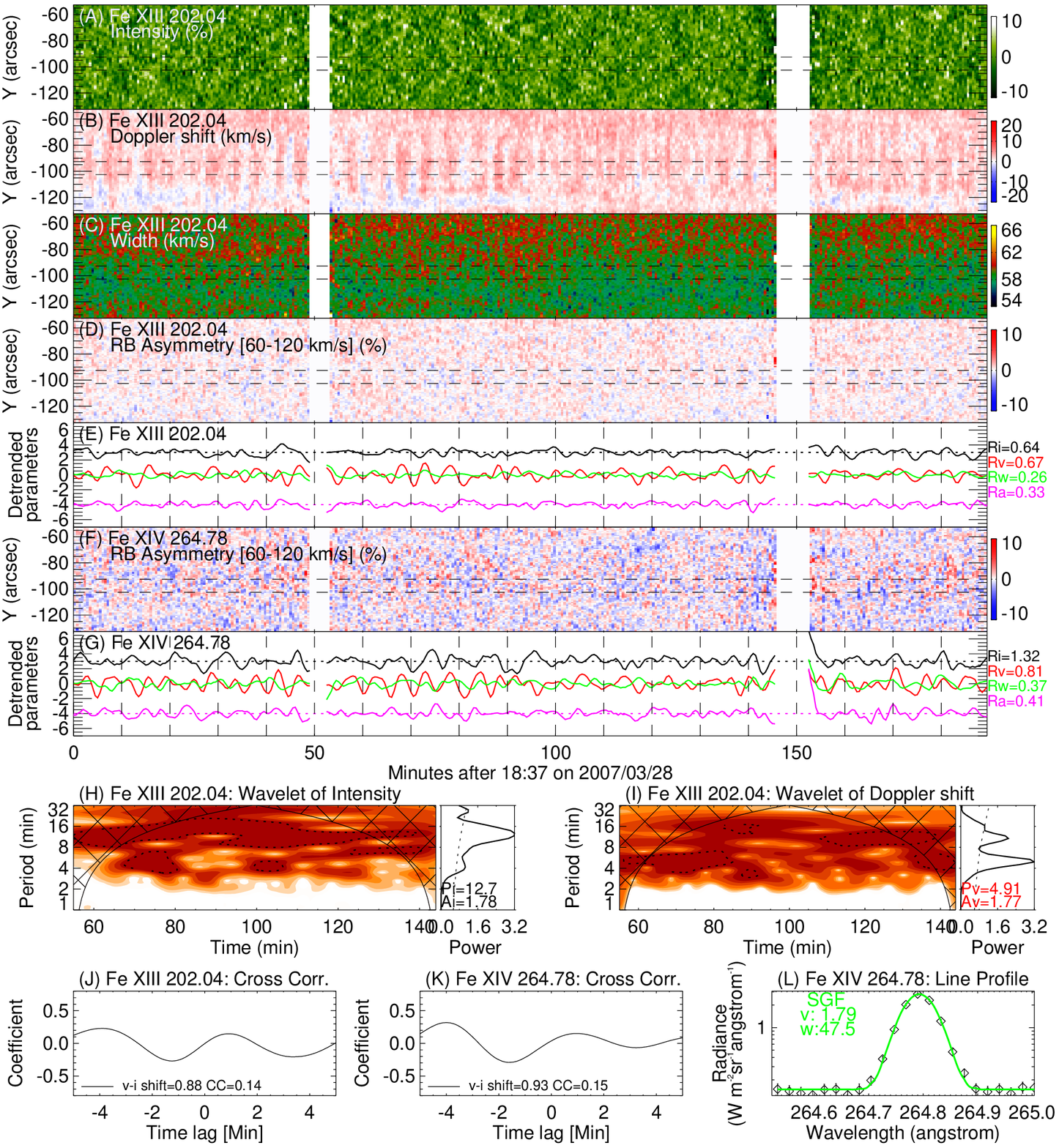}} \caption{ Similar to Figure~\ref{fig.3} but for the range of
y=-133$^{\prime\prime}$$\sim$-53$^{\prime\prime}$ in Figure~\ref{fig.1}(B). In panels (J)-(K) only the results of cross correlation between 
the Doppler shifts and intensities are shown. In panel (L) only the single Gaussian fit is applied to the observed line profile.  } \label{fig.5}
\end{figure*}

The temporal evolution of all of the four line parameters of Fe~{\sc{xiii}}~202.04\AA{} is detailed in Figure~\ref{fig.5}(A)-(D). Clearly,
oscillation 2 is mainly present in the Doppler shift. No clear discernible variations are found in the line width and RB asymmetry. The Fe~{\sc{xiv}}~264.78\AA{} line shows behaviors similar to Fe~{\sc{xiii}}~202.04\AA{}, as can be seen from panels (F) and (G). The line profiles in this loop top region are typically symmetric and an example is presented in panel (L). The variation of the detrended intensity is smaller than that of oscillation 1. But from panel (G) we see a seemingly correspondence between the intensity and Doppler shift variations. The correspondence can also be seen from Figure~\ref{fig.4} and is more clear in hotter lines such as Fe~{\sc{xiv}}~274.20\AA{} and Fe~{\sc{xv}}~284.16\AA{}. 
From the analysis of cross correlation, we find a maximum correlation coefficient at the time lag of about 0.9 min between the time series of intensity and Doppler shift for the two lines shown in Figure~\ref{fig.5}. Such a time lag corresponds to a phase shift of about $\pi$/2 or 1/4 period, given a Doppler shift oscillation period of 4.91$\pm$0.67 min. However, since the correlation coefficient is smaller than 0.3 (which can easily be caused by random, chance correlations) and the dominant period (12.7 min$\pm$1.01) of the intensity is largely different from that of the Doppler shift, we can not make any solid conclusion about it. The oscillation amplitude is 1.78\% for the intensity and 1.77 km~s$^{-1}$ for the Doppler shift. Clearly, the intensity fluctuation is much smaller than that of oscillation 1. Whilst the amplitude of the Doppler shift oscillation is very similar to that of oscillation 1.

\section{Statistical results and discussion}
\begin{table*}[]
\caption[]{Persistent Doppler shift oscillation events observed by EIS from Feb to April in 2007. The following information is listed for each event: observation date (yyyymmdd), starting time (hhmmss), lowest y-pixel of the selected 11-pixel region on the slit (Y), root-mean-square values of the detrended intensity (Ri, \%), Doppler shift (Rv, km~s$^{-1}$), line width (Rw, km~s$^{-1}$) and RB asymmetry (Ra, \%), intensity oscillation period and uncertainty (Pi \& Piun, min), Doppler shift oscillation period and uncertainty (Pv \& Pvun, min), amplitudes of intensity (Ai, \%) and Doppler shift (Av, km~s$^{-1}$) oscillations, electron density and uncertainty (N \& Nerr, log cm$^{-3}$), oscillation type and location.} \label{tab.1}
\begin{center}
\begin{tabular}{p{0.1cm}  p{1.2cm}  p{0.8cm}  p{0.4cm}  p{0.4cm}  p{0.4cm}  p{0.4cm}  p{0.4cm}  p{0.4cm}  p{0.4cm}  p{0.4cm}  p{0.5cm}  p{0.4cm}  p{0.4cm}  p{0.4cm}  p{0.4cm}  p{2.2cm}}
\hline id & date & time & Y &  Ri &  Rv &  Rw &  Ra &  Pi &  Piun &  Pv &  Pvun & Ai & Av & N & Nerr & Type, location \\
\hline 
 1 &20070201 &013212 &351 & 1.07 & 0.44 & 0.30 & 0.46 &10.7 & 2.44 & 9.82 & 2.05 & 3.89 & 1.65 & 9.23 & 0.04 &    I, loop leg \\
 2 &20070202 &004912 &245 & 0.89 & 0.49 & 0.51 & 0.65 & 7.57 & 3.19 &10.7 & 4.79 & 2.36 & 1.45 & 9.25 & 0.04 &    I, loop leg \\
 3 &20070203 &005642 &193 & 1.14 & 0.59 & 0.56 & 0.63 & 8.77 & 3.00 & 7.57 & 2.44 & 2.71 & 1.60 & 9.22 & 0.04 &    I, loop leg \\
 4 &20070220 &175013 &296 & 4.57 & 2.79 & 2.33 & 2.60 &11.6 & 2.04 &12.7 & 2.45 &14.0 & 8.47 & 9.65 & 0.13 &    I, loop leg \\
 5 &20070221 &021812 &310 & 3.45 & 1.62 & 1.33 & 3.12 &10.7 & 2.90 & 9.82 & 2.05 &11.8 & 5.19 & 9.32 & 0.17 &    I, loop leg \\
 6 &20070326 &150012 &168 & 0.87 & 0.48 & 0.48 & 0.51 & 8.26 & 1.88 & 9.82 & 2.05 & 2.53 & 1.45 & 9.21 & 0.05 &    I, loop leg \\
 7 &20070327 &143412 &220 & 1.24 & 0.56 & 0.41 & 0.49 &13.8 & 2.42 & 9.82 & 4.89 & 4.11 & 1.40 & 9.65 & 0.10 &    I, loop leg \\
 8 &20070328 &031531 &264 & 1.84 & 0.89 & 0.77 & 0.81 &13.8 & 3.17 &11.7 & 2.99 & 5.50 & 2.63 & 9.03 & 0.05 &    I, loop leg \\
9 &20070328 &183726 &303 & 2.08 & 0.56 & 0.77 & 0.76 & 9.01 & 2.64 &10.7 & 2.70 & 6.25 & 1.67 & 9.52 & 0.07 &    I, loop leg \\
 10 &20070216 &163920 &345 & 1.93 & 0.70 & 0.46 & 0.71 & 5.35 & 1.29 & 4.13 & 1.46 & 4.15 & 1.43 & 9.67 & 0.07 &      II, QS BP \\
 11 &20070221 &194813 &394 & 1.22 & 0.54 & 0.34 & 0.56 &10.7 & 1.34 & 3.47 & 0.43 & 4.73 & 1.57 & 8.83 & 0.05 &   II, loop top \\
 12 &20070223 &125943 &384 & 1.09 & 0.46 & 0.22 & 0.45 & 6.37 & 1.46 & 4.91 & 2.19 & 2.09 & 0.95 & 9.27 & 0.07 &   II, loop top \\
 13 &20070223 &185342 &380 & 1.06 & 0.56 & 0.35 & 0.55 & 6.14 & 0.87 & 3.47 & 0.61 & 2.38 & 1.15 & 9.25 & 0.08 &   II, loop top \\
 14 &20070224 &005742 &381 & 1.27 & 0.54 & 0.34 & 0.57 & 5.84 & 1.68 & 3.18 & 0.67 & 2.48 & 1.08 & 9.31 & 0.09 &   II, loop top \\
15 &20070224 &130643 &370 & 0.88 & 0.63 & 0.25 & 0.31 & 4.50 & 1.03 & 3.47 & 0.30 & 1.67 & 1.76 & 9.32 & 0.09 &   II, loop top \\
16 &20070225 &112113 &185 & 0.91 & 0.66 & 0.29 & 0.46 & 7.57 & 3.06 & 4.91 & 1.29 & 1.95 & 1.58 & 9.10 & 0.05 &    II, unclear \\
17 &20070326 &150012 & 29 & 0.79 & 0.68 & 0.27 & 0.38 & 8.26 & 2.92 & 5.35 & 0.93 & 2.21 & 2.13 & 9.00 & 0.04 &   II, loop top \\
18 &20070327 &160926 &247 & 0.80 & 0.74 & 0.28 & 0.38 & 6.94 & 2.05 & 4.91 & 1.12 & 2.04 & 1.70 & 9.16 & 0.03 &   II, loop top \\
19 &20070328 &031531 & 12 & 0.84 & 0.72 & 0.27 & 0.46 & 8.26 & 3.44 & 7.57 & 2.59 & 2.03 & 1.76 & 8.94 & 0.04 &   II, loop top \\
20 &20070328 &183726 & 40 & 0.64 & 0.67 & 0.26 & 0.33 &12.7 & 1.01 & 4.91 & 0.67 & 1.78 & 1.77 & 9.14 & 0.03 &   II, loop top \\
21 &20070419 &191102 &189 & 3.17 & 0.97 & 0.44 & 0.65 & 6.94 & 1.02 & 5.35 & 1.29 & 8.39 & 2.73 & 9.08 & 0.03 &      II, QS BP \\
\hline
\end{tabular}
\end{center}
\end{table*}

Table~\pref{tab.1} lists the observational information and analysis results, including observation time, locations on the slit, root-mean-square
values of all line parameters, oscillation amplitudes, periods and uncertainties of the intensity and Doppler shift, electron densities, and locations,
for all oscillation events we identified. These parameters were derived from an averaged time series over 11 pixels along the slit in each event. We carefully selected these 11 Y-pixels in each event and made sure that 1) the temporal behavior at these 11 pixels is almost the same, 2) this 11-pixel region covers the major oscillating part. Similarly, we detrended each time series by subtracting a 10-minute running average. We mainly used the strong Fe~{\sc{xiii}}~202.04\AA{}
line which is available in all observations for our statistical studies. The weaker but clean Fe~{\sc{xiv}}~264.78\AA{} line was also analyzed for a confirmation of the characteristics revealed by Fe~{\sc{xiii}}~202.04\AA{}. The line pair of Fe~{\sc{xii}}~196.64/195.12\AA{} or Fe~{\sc{xii}}~186.88/195.12\AA{} was used to diagnose the electron densities. Images obtained by the Transition Region and Coronal Explorer \citep[TRACE,][]{Handy1999}, the X-Ray Telescope \citep[XRT,][]{Golub2007} and the Extreme Ultraviolet Imaging Telescope \citep[EIT,][]{Delaboudiniere1995} were used for identifying the coronal structures associating with these oscillations. Based on these statistics we can subdivide the oscillations we found here into two types. One type (type I) are mainly found in the lower part of AR loops (Sect 4.1) and they are probably related to recurring upflows. The other type of oscillations (type II) are found mostly near the top of loops, and the observational facts seem to suggest an interpretation of transverse wave. We have to mention that event 6 is very close to a localized bright region (TRACE 195\AA{} passband) in the core of an AR. But from the XRT image it is clear that the oscillating part (the localized bright region in TRACE 195\AA{} passband) is the leg (or lower part) of a hot loop system. We could not find any context images for event 16 so that its location is unclear.

\subsection{Type I oscillations: recurring upflows?}
\begin{figure*}
\centering {\includegraphics[width=0.9\textwidth]{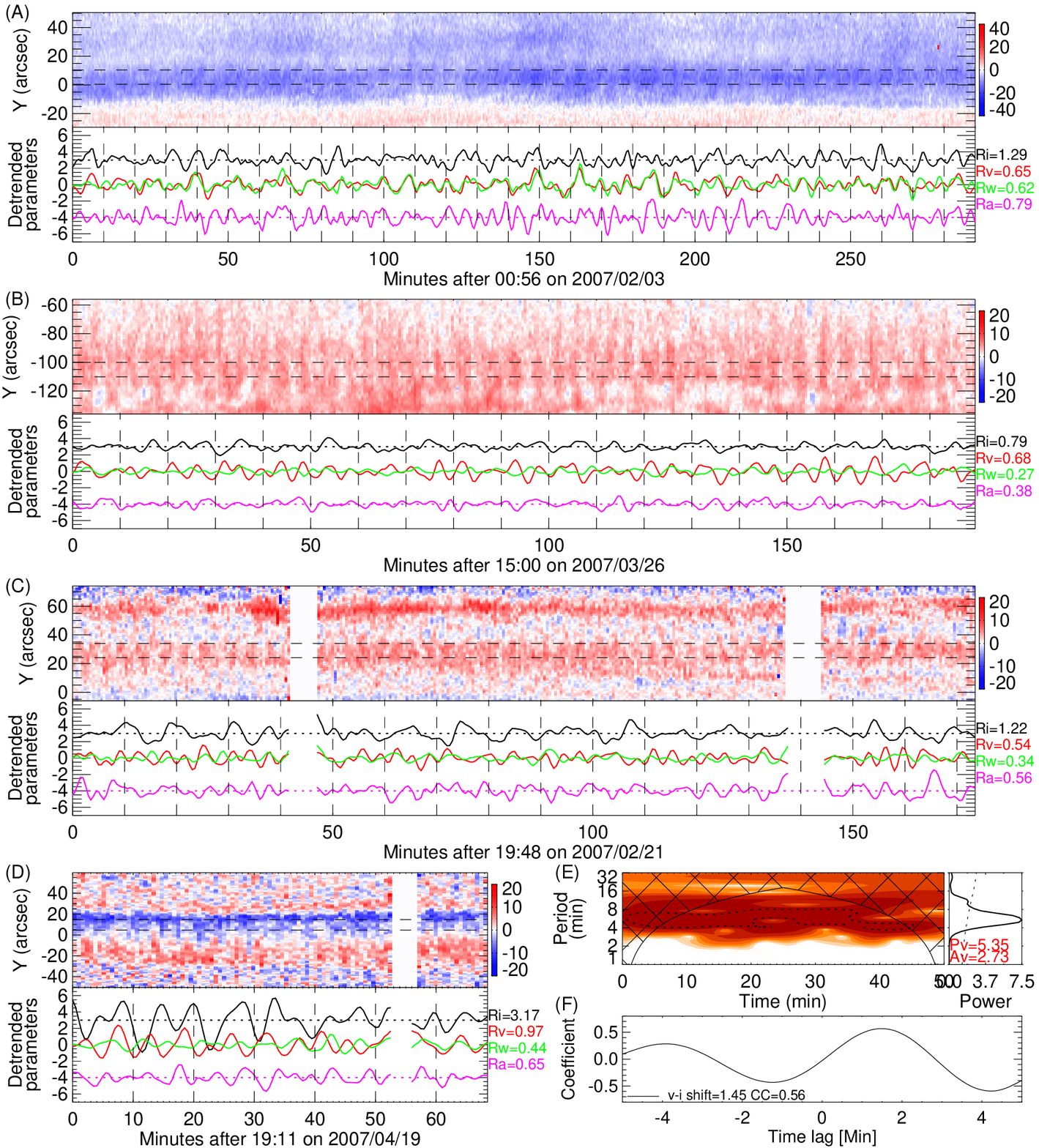}} \caption{ (A) The Doppler shift and time series of all line parameters of Fe~{\sc{xiv}}~264.78\AA{} 
for a type I oscillation event. The line styles, units, and illustration methods are the same as in Figure~\ref{fig.3}(G). (B)-(D): The Doppler shift and time series of all line parameters of Fe~{\sc{xiii}}~202.04\AA{} for three type II oscillation events. The line styles, units, and illustration methods are the same as in Figure~\ref{fig.5}(E). (E)-(F) Wavelet (of Doppler shift, similar to Figure~\ref{fig.5}(I)) and cross correlations analysis (similar to Figure~\ref{fig.5}(J)) for the event shown in panel (D).} \label{fig.6}
\end{figure*}

We found nine oscillation events which reveal properties similar to oscillation 1 discussed in the previous section. One example is shown in
Figure~\ref{fig.6}(A). These oscillations are all found at the foot point regions (lower parts) of AR
loops. They are characterized by coherent behaviors of all line parameters (line intensity, Doppler shift, line width and profile asymmetry),
obvious blue shift and blueward asymmetry throughout almost the entire observational duration. The time lag between the Doppler shift and other parameters is very close to zero and their correlation coefficients are typically larger than 0.3 for all of these events. The line widths are obviously enhanced at the
oscillation locations. These oscillations are named type I oscillations in the following discussions.

For these type I oscillations, the global wavelet power is often distributed over a wide range of periods. The dominant periods (period corresponding to the peak of a global wavelet) range from 7.57-13.8 min for the intensity and 7.57-12.7 min for the Doppler shift, with an average of 10.5$\pm$2.3 min for the intensity and 10.3$\pm$1.4 min for the Doppler shift. The oscillation amplitudes are in the range 2.36-14.0\% for the intensity and 1.40-8.47 km~s$^{-1}$ for the Doppler shift. The average amplitudes of the
intensity and Doppler shift are 5.9\% and 2.8 km~s$^{-1}$, respectively. In two events starting at 17:50:13 on Feb 20 (previously analyzed by \cite{Tian2011a} and \cite{Nishizuka2011}) and 02:18:12 on Feb 21 we see an intensity amplitude of about 13\% and Doppler shift amplitude of about 7 km~s$^{-1}$. The root mean square values of the line width and RB asymmetry for these two events are about 2 km~s$^{-1}$ and 3\%, respectively. These values are much larger than those of other events. The magnitudes of fluctuations are likely influenced by effects such as LOS projection and the foreground/background emission. 

The line profiles are usually enhanced in the blue wings, suggesting the existence of at least two emission components. Compared to the single Gaussian fit, the RB-guided double Gaussian fit \citep{DePontieu2010,Tian2011a,Tian2011c,Tian2012} usually does a much better job. From
Figure~\ref{fig.3}(L) one can also conclude that the enhanced line width and significant blue shift derived from the single Gaussian fit are
actually caused at least partly by the superposition of a weak high-speed upflow component on a strong background emission component
\citep{Tian2011c}.

\begin{figure*}
\centering {\includegraphics[width=0.9\textwidth]{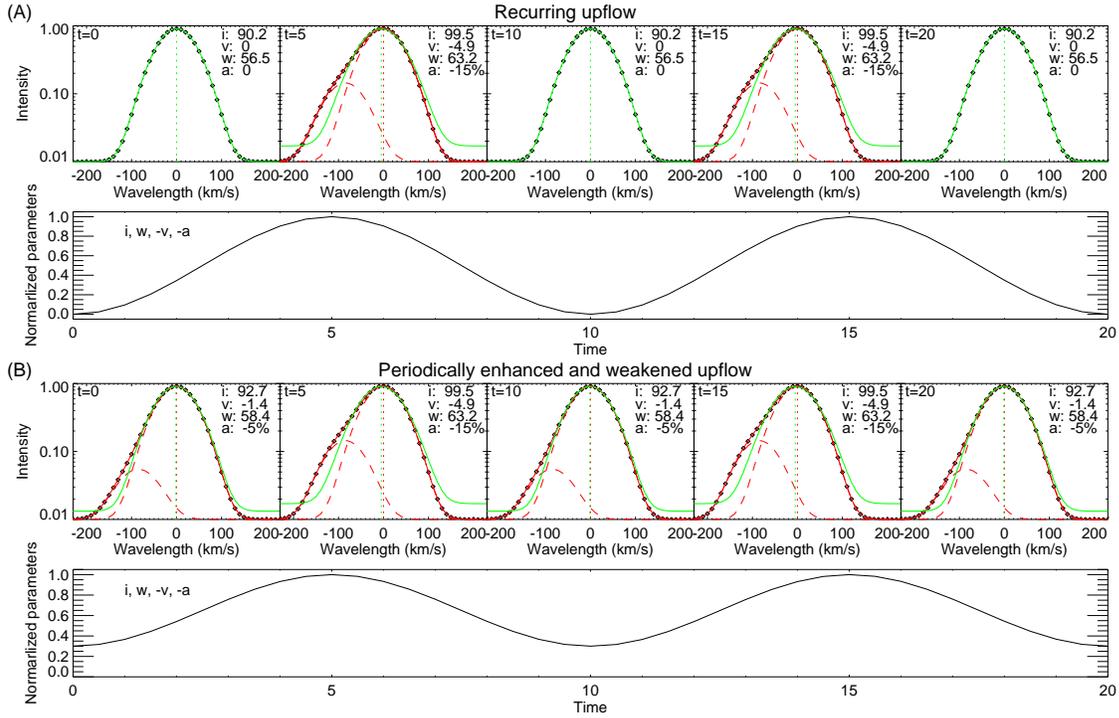}} \caption{ Synthetic line profiles (diamonds) and time series of the line parameters in
the scenarios of recurring upflows (A) and periodically enhanced and weakened upflows (B). In the panels of line profiles, the green/red solid
line is the single/double Gaussian fit. The two dashed red lines represent the primary (background) and secondary (upflow) components,
respectively. The red and green vertical lines indicate the line center of the primary component and single Gaussian fit, respectively. The
single Gaussian fit parameters (i-line intensity, v-Doppler shift, w-line width) and RB asymmetry (a) are shown in each panel. } \label{fig.7}
\end{figure*}

These observational facts point to the explanation of recurring high-speed upflows (or quasi-periodically enhanced and weakened upflows) for
type I oscillations, as previously suggested by \cite{DePontieu2010} and \cite{Tian2011a}. Figure~\ref{fig.7} presents synthetic line profiles and time series of the line parameters in the scenarios of recurring upflows and periodically enhanced and weakened upflows. The background coronal emission component and high-speed upflow component are represented by the two red dashed lines. Clearly, continuous high-speed upflows with periodic enhancement of the flow intensity can easily explain the quasi-periodic enhancement of the line intensity, line width, blue shift and blueward asymmetry. In real observations, the flow intensity might be enhanced differently in different periods \citep[clearly seen in][]{Tian2011a}, which can sometimes cause a deviation of the dominant period derived from wavelet analysis from the real period of recurrence. 

Correlated changes between intensity and Doppler shift are usually interpreted as propagating slow mode magnetoacoustic waves \citep[e.g.,][]{Wang2009a,Wang2009b,Nishizuka2011}. However, here we can see that such a correlation also exists in the case of recurring upflow. It seems that the line width and asymmetry are the key to distinguish between slow waves and upflows. \cite{Verwichte2010a} argued that slow waves can also cause line asymmetries when the emission line is averaged over an oscillation period or when a quasi-static plasma component in the line of sight is included. However, the frequency doubling of the line width, as shown in their Figure~3, is not seen in our type I oscillations. In addition, the irregular quasi-periodic changes (change differently in different periods) of the intensity and Doppler shift are also expected in the scenario of recurring upflows since each repetitive upflow is likely to be independent of each other. So at present we conclude that the type I oscillations we identified here are more likely to be dominated by quasi-periodically enhanced and weakened upflows. \cite{Heggland2009} and \cite{McLaughlin2012} have recently demonstrated that such quasi-periodic outflows can be generated by oscillatory reconnection \citep{McLaughlin2009}. \cite{Morton2012} also proposed that the recurrent plasma ejections they observed are reconnection-driven and similar to type-II spicules. Based on an observation of a supersonic blob, \cite{Srivastava2012} suggested that multi-temperature plasma blobs could be generated by a recurrent three-dimensional (3-D) reconnection process via the separator dome below the magnetic null point, between the emerging flux and pre-existing field lines in the lower solar atmosphere. Indeed, quasi-periodic flow pulses associated with a null point have recently been reported \citep{Su2012}. On the other hand, \cite{Tripathi2012} explained the hot plasma upflows observed in the warm loops as evidence of chromospheric evaporation in quasi-static coronal loops. Whilst in the 3-D simulation of \cite{Martnez-Sykora2011}, the chromospheric plasma is ejected as result of being squeezed by the magnetic tension.

We would like to point out that slow waves may also exist. The disappearance of enhanced line width and blueward asymmetry at higher parts of loops might be caused by LOS effect or low signal to noise ratio of EIS. But it might also be explained by slow waves \citep{Nishizuka2011}. \cite{Ofman2012} recently found that periodically driven upflows can excite undamped slow waves in loops. They showed that the properties of quasi-periodic flows are manly related to driving source, while the properties of waves are determined by the structures of loops. It is likely that high-speed upflows dominate in footpoints (lower corona) of some loops and slow waves dominate higher up in the loops. More recently, through a detailed analysis of the temperature dependence of the speeds of the propagating disturbances, \cite{Kiddie2012} found signatures of both upflows and slow waves.  On the other hand, slow waves associated with the leakage of p-modes may well exist in the chromosphere \citep[e.g.,][]{DeWijn2009} and some of them might propagate further up into the corona and mix with the upflows. 

These high-speed upflows in lower parts of loops may play an important role in the supply of mass and energy to the hot corona and probably also the solar wind \citep{DePontieu2009,DePontieu2011}. There are suggestions that the blue shifts of the order of 30~km~s$^{-1}$ at boundaries of some ARs
\citep{Marsch2004,Marsch2008,DelZanna2008,Tripathi2009,Murray2010,Warren2011,Young2012}, as derived from a single Gaussian fit, are signatures of the
nascent slow solar wind \citep{Sakao2007,Harra2008,Doschek2008,He2010,Brooks2011,Slemzin2012,Zangrilli2012}. This might be true for the upflows along open field lines originating from the AR boundaries. But we have to point out that the speeds of these outflows are around 100 km~s$^{-1}$ instead of 30~km~s$^{-1}$, as
revealed by the RB asymmetry analysis and double Gaussian fit
\citep{Hara2008,DePontieu2009,DePontieu2010,McIntosh2009a,McIntosh2009b,Peter2010,Bryans2010,Martnez-Sykora2011,Dolla2011,Ugarte-Urra2011,Tian2011a,Tian2011c,Tian2012}. In addition, some of these outflows are clearly associated with legs of closed field lines \citep{Tian2011c,Boutry2012} and they are followed by cooling downflows \citep{Ugarte-Urra2011,Kamio2011,McIntosh2012} instead of propagating into the interplanetary space. These outflows may be the coronal counterpart of type-II spicules which could supply heated mass to the corona \citep{DePontieu2011}. Taking an average speed of 100 km~s$^{-1}$ and assuming a density of log ({\it N}$_{e}$/cm$^{-3}$)=8, the mass flux density of these outflows is estimated to be about 3.3$\times$10$^{-10}$ g~cm$^{-2}$~s$^{-1}$ if we use a temporal filling factor of 0.2. 

\subsection{Type II oscillations: waves?}
Twelve oscillation events are found to reveal characteristics different from those of type I oscillations. These oscillations are named type II oscillations in the following discussions. These oscillations are found to be associated with the upper parts or tops of AR loops. Two of them are actually found in quiet-Sun bright points, which are also believed to consist of a group of miniature loops. The oscillation is most prominent in Doppler shift and sometimes the cross correlation analysis reveals a $\pi$/2 phase shift between the intensity and Doppler shift oscillations. The variations of the line intensity, line width and RB asymmetry are all typically smaller than those of type I oscillations. The average root-mean-square values of the detrended line intensity/line width/RB asymmetry are 1.9\%/0.83 km~s$^{-1}$/1.11\% for type I and 1.2\%/0.31 km~s$^{-1}$/0.48\% for type II oscillations.  

\begin{figure*}
\centering {\includegraphics[width=0.8\textwidth]{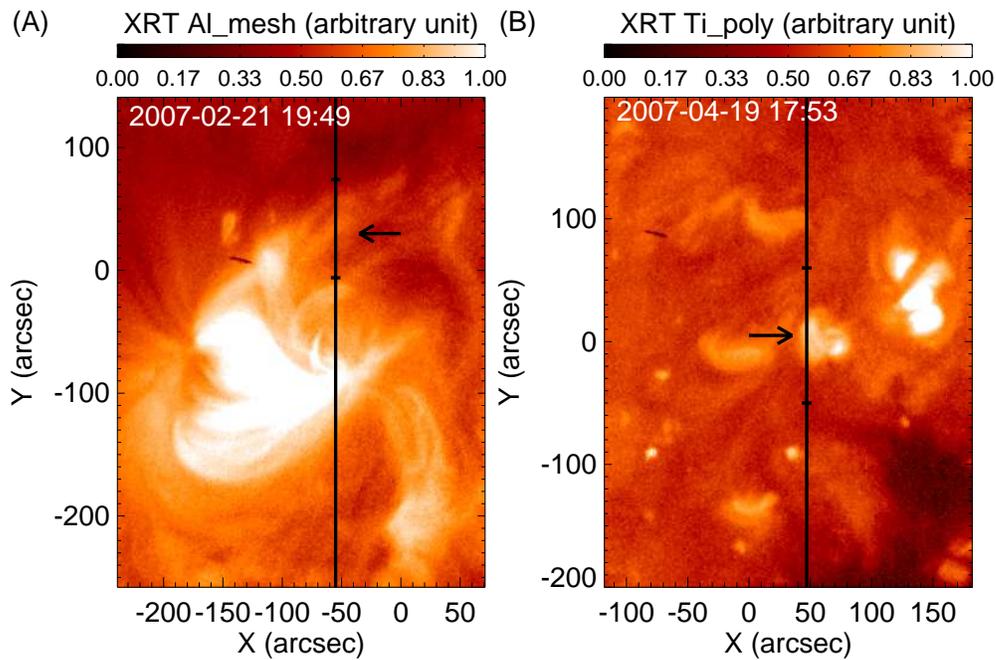}} \caption{ Coronal structures associated with two type-II oscillation events (events 11 \& 21). (A) An XRT image taken at 19:49 on 2007 Feb 21. The black vertical line indicates the location of the EIS slit. The region between the two horizontal bars corresponds to the y-range of the Dopplergram in Figure~\ref{fig.6}(C). The arrow points the approximate location of oscillation event 11. (B) An XRT image taken at 17:53 on 2007 April 19. The region between the two horizontal bars corresponds to the y-range of the Dopplergram in Figure~\ref{fig.6}(D). The arrow points the approximate location of oscillation event 21.}
\label{fig.8}
\end{figure*}

One type II oscillation event (event 17 in Table~\pref{tab.1}) presented in Figure~\ref{fig.6}(B) is characterized by almost vertical stripes of enhanced red shift. In addition, the slit is found to be almost parallel to the loop plane and covered at least 50$^{\prime\prime}$ (coherence scale) of the upper segment of the loop. Such characteristics are very similar to the events 20 (see Figure~\ref{fig.1} \& Figure~\ref{fig.5}) and 19 (not shown here) and the coherence scales are around 50$^{\prime\prime}$. There are signatures of correspondence between the intensity and Doppler shift variations. The time lag between the two suggests a possible $\pi$/2 phase shift, although sometimes the correlation coefficient is smaller than 0.3 and the dominant period of intensity is larger than that of the Doppler shift. In all of these events, we do not see clear signatures of propagating features (inclined stripes) in the time-slit diagrams of intensity and Doppler shift. Such an observational fact suggests that the phase speed is at least of the order of 1200 km~s$^{-1}$ (50$^{\prime\prime}$ divided by the exposure time 30 s) if these oscillations are signatures of propagating features. Such a high speed is much larger than the sound speed and comparable to the Alfv\'en speed in the corona.

The event 11 (presented in Figure~\ref{fig.6}(C)) only exhibits oscillation in a narrow region ($\sim$10$^{\prime\prime}$) on the slit, which might suggest that the loop width is about 10$^{\prime\prime}$ since the slit was found to make a large angle with respect to the loop top (see Figure~\ref{fig.8}(A)). No clear correlation was found for the intensity and Doppler shift variations in this event. We have found that the top parts of loops associated with oscillating events 12, 13, 14, 15 and 18 all made a relatively large angle with respect to the EIS slit. The coherence scales of these oscillations are typically 10$^{\prime\prime}$-30$^{\prime\prime}$. These type II oscillations are usually associated with quasi-periodic appearance of (enhanced) red shift and no significant blue shift has been found, which is a puzzle but might be partly related to the assumption of zero Doppler shift of the average line profile. 

Most of these type II oscillations are found in ARs. However, two events (events 10 \& 21) are found to be associated with coronal bright points (BPs). In these two events we see a relatively large intensity fluctuation and a clear $\pi$/2 phase shift between the intensity and Doppler shift oscillations. Analysis results of event 21 are presented in Figure~\ref{fig.6}(D)-(F) and the oscillating BP can be found in Figure~\ref{fig.8}(B). The blue shift between y=-20$^{\prime\prime}$$\sim$5$^{\prime\prime}$ and red shift between y=-45$^{\prime\prime}$$\sim$-20$^{\prime\prime}$ might be related to our assumption of zero Doppler shift of the average line profile. It could also be caused by a siphon flow along the small BP loop system. 

The periods of the type II Doppler shift oscillations are in the range of 3-6 min, with a mean of 4.6 min. The global wavelets usually peak sharply and thus the uncertainties of the periods are smaller compared to type I oscillations. The average oscillation amplitude, as computed from the global wavelet, is 1.6 km~s$^{-1}$. Such a small oscillation amplitude is extremely difficult to be identified from on-disk imaging observations if they correspond to transverse displacement oscillations. With spectroscopic observations, we can clearly see them. 

Unlike type I oscillations, it is difficult to find any significant profile asymmetry at the oscillation locations. The almost symmetric line profiles seem to exclude the presence of high-speed outflows which are believed to be responsible for the type I oscillations. Perhaps a possible mechanism for these type II oscillations is the intermittent obscuration by fibrils \citep{DePontieu2005,DeWijn2007}, or the coronal response of fibril motion. Imagining cool chromospheric fibrils going up and down quasi-periodically, the overlying hot coronal plasma can be forced upward and released downward with the same period. Such a scenario seems to be similar to vertical piston motion. In this case the entire coronal plasma is moving at a small speed, which may explain the absence of profile asymmetry and line broadening. However, it is unknown whether the fibril motion can affect the tops of these coronal loops which are often high (e.g., $\sim$50$^{\prime\prime}$ for the type I oscillation event shown in Figure~\ref{fig.1}) in the corona.  

\subsubsection{Mode identification and coronal seismology}
More likely, the type II oscillations are signatures of MHD waves. In the following we discuss possible wave modes which might explain our observations and their roles in coronal seismology when applicable. 

\subsubsubsection{Propagating slow wave}
Slow mode magneto-acoustic waves are compressible modes and a correlation (or anti-correlation) between intensity and Doppler shift is expected in the propagating slow wave. Our observations reveal very small intensity fluctuations of all type II oscillation events except events 10 and 21. Moreover, we do not see any clear signature of correlation (or anti-correlation) between intensity and Doppler shift oscillations in any of our type II oscillation events. Thus, we can rule out the possibility of propagating slow waves. 

The top parts of oscillating loops are more or less parallel to the EIS slit in events 17, 19 and 20 (see the example in Figure~\ref{fig.1}). Given a typical coronal sound speed of 200 km~s$^{-1}$, the wave would need 180 s (much larger than the cadence of 30 s) to propagate through a distance of 50$^{\prime\prime}$ so that we should be able to see propagating features (inclined stripes) in the time-slit diagrams of intensity and Doppler shift. The absence of such propagating features again rules out the possibility of propagating slow waves.

\subsubsubsection{Standing slow wave}
The seemingly $\pi$/2 phase shift between intensity and Doppler shift in some oscillation events suggests the possibility of standing slow waves \citep{Sakurai2002,Kitagawa2010}. For the fundamental mode the wave period is determined by twice the loop length divided by sound speed \citep{Wang2002}. Due to the strong foreground emission and the mixture with emission from surrounding structures, it is often difficult to accurately locate the legs of the oscillating loops (so loop reconstruction techniques are highly desired, e.g., \cite{Feng2007} \& \cite{Syntelis2012}). Supposing that the oscillating loop legs are located at [300$^{\prime\prime}$,-130$^{\prime\prime}$] and [320$^{\prime\prime}$,-30$^{\prime\prime}$] in Figure~\ref{fig.1} and assuming a semi-circular loop, we obtain a loop length of 160$^{\prime\prime}$. Using a sound speed of 200 km~s$^{-1}$, the period was estimated to be 19.22 min, which is much larger than the observed period (4.91 min) of event 20. Therefore, fundamental mode of standing slow waves can not explain type II oscillations in such large loops. A possibility that can be considered is a higher harmonic. But compared to fundamental mode, harmonic modes are difficult to be excited in coronal loops. In addition, for most of the type-II oscillations the correlation between intensity and Doppler shift is poor (coefficient smaller than 0.3), which again does not favor a slow wave interpretation. 

However, standing slow waves may not be excluded for oscillations found in smaller loops such as the events of Feb 16 and April 19 (events 10 \& 21 in Table~\pref{tab.1}, Figure~\ref{fig.6}(D)-(F)). In these two events we also see relatively larger fluctuation of intensity, similar periods of intensity and Doppler shift oscillations, and a $\pi$/2 phase shift between intensity and Doppler shift, which favors the interpretation of standing slow waves if the intensity fluctuations are signatures of density change \citep[e.g.,][]{Sakurai2002,Kitagawa2010}. In such small BP loop systems, undamped propagating waves from one loop leg might reflect when they reach the other loop leg, thus producing standing waves. The first oscillation in \cite{Erdelyi2008} might be of similar type. However, as we will explain in the following, the intensity fluctuations might be related to the fact that different parts of a loop are sampled at different times. In this case, the intensity fluctuations are not related to density change but may be caused by loop displacement during transverse oscillations. 

\subsubsubsection{Standing kink wave}
For the fundamental mode of a standing kink wave the period is determined by twice the loop length divided by the kink speed \citep[e.g.,][]{Roberts1984,Wang2002,Aschwanden2011} defined as:

\begin{equation}
\emph{$c_k=\sqrt{\frac{2}{1+\rho_e/\rho_i}}V_{Ai}$}\label{equation1},
\end{equation}
 
\noindent where $\rho_i$ and $\rho_e$ are internal and external densities, respectively. $V_{Ai}$ is the internal Alfv\'en speed which is related to the internal magnetic field ($B_i$) and density in the form of:

\begin{equation}
\emph{$V_{Ai}=\frac{B_i}{\sqrt{4\pi \rho_i}}$}\label{equation2},
\end{equation}

\noindent In order to explain the period (4.91 min) of oscillation event 20, the kink speed needs to be about 775 km~s$^{-1}$. This value is in the range (543-2322 km~s$^{-1}$) of those measured from TRACE transverse oscillations \citep{Ofman2002}. However, we have to bear in mind that the value of $c_k$ depends on the magnetic field strength and density contrast between the loop and background, both of which can vary a lot in different cases. In addition, the oscillation could be modulated by some continuous driver like p-modes, as suggested by the fact that the oscillations are persistent and undamped. Thus, the period may differ from the value predicted by twice the loop length divided by the kink speed. Also, the above phase speed is based on cylindrical or slab geometry approximation of real coronal loops. Finally, kink modes can also exhibit some intensity disturbances if the angle between the LOS and the oscillating structure is not exactly 90 degree \citep{Cooper2003,Wang2012b}. So the standing kink mode can not be ruled out. The weakly damped Doppler shift oscillation event (2nd event) interpreted as kink mode by \cite{Erdelyi2008} may be of similar type. They did not find any flare or prominence eruption associated with the oscillation, which is also similar to our type II oscillations.

If these waves are standing kink waves, we can in principle obtain the internal Alfv\'en speed and magnetic field strength according to Equations~\pref{equation1}\&\pref{equation2}. Using the line ratio method, we have calculated the internal and external densities from observations. However, the density ratio $\rho_e/\rho_i$ we obtained is usually in the range of 0.2-0.7, which is much larger than the density ratio normally assumed ($\sim$0.08). Such results can be explained by the fact that LOS summation substantially reduces the density contrast \citep{DeMoortel2012}. Fortunately, the density enters through square root and the dependence of the phase speed on the density and the contrast is weak. In fact there is only about 25\% difference in $c_k$ between the density ratios of 0.08 and 0.7. Here we simply took the typical value of 0.08 \citep{Aschwanden2011} and obtained a value of 570 km~s$^{-1}$ for the internal Alfv\'en speed in oscillation event 20. Taking the calculated density of log ({\it N}$_{i}$/cm$^{-3}$)=9.14, the magnetic field strength of the oscillating loop is estimated to be 11.6 G.  Here the internal mass density $\rho_i$ is calculated as 1.2$m_p${\it N}$_{i}$, where $m_p$ is the proton mass and the constant of 1.2 is due to the consideration of Hellium abundance \cite[e.g.,][]{Priest1982,Wang2009a}. Using a density uncertainty of log ({\it N}$_{i}$/cm$^{-3}$)=0.03, a period uncertainty of 0.67 min, and assuming a 20\% uncertainty of the loop length, the uncertainty of the estimated magnetic field \citep{Nakariakov2001} would be $\sim$25\% (2.9 G). Note that Equation~\pref{equation1} is based on the simple configuration of a thin and straight cylindrical tube embedded in a uniform magnetic atmosphere \citep{Edwin1983,Roberts2008}, which might be too idealized for our coronal loop oscillations. In addition, through numerical simulations \cite{DeMoortel2009b} have demonstrated that the magnetic field derived from coronal seismology might differ from the real magnetic field by $\sim$50\% in some cases.

\subsubsubsection{Propagating Alfv\'enic wave}
Ubiquitous coronal Alfv\'en waves have been found by observations of both the Coronal Multi-channel Polarimeter \citep[CoMP,][]{Tomczyk2007,Tomczyk2008,Tomczyk2009} and the Atmospheric
Imaging Assembly \citep[AIA,][]{Lemen2012} on board the Solar Dynamics Observatory (SDO) \citep{McIntosh2011}. \cite{VanDoorsselaere2008} proposed that these waves are better called fast mode kink waves rather than Alfv\'en waves. However, another theoretical paper \cite{Goossens2012} presents a different view and clearly states that these waves can be considered as surface Alfv\'en waves or Alfv\'enic waves in the nomenclature of \cite{Goossens2009}.  \cite{McIntosh2011} also pointed out in the online supplementary information that: 1) in the highly dynamic atmosphere there are no stable cylindrical waveguides with straight magnetic field lines that \cite{VanDoorsselaere2008} require for kink waves; 2) the term Alfv\'en wave is commonly used by the communities of fusion plasma physics and space physics for a largely incompressible transverse wave for which the major restoring force is the magnetic tension. Here we follow \cite{Goossens2009} and \cite{McIntosh2011} and use the term Alfv\'enic waves to describe more loosely the magnetic oscillations we observed with EIS. Here we follow \cite{Goossens2009} and \cite{McIntosh2011} and use the term Alfv\'enic waves to describe more loosely the magnetic oscillations we observed with EIS.

 \cite{Tomczyk2007} and \cite{Tomczyk2009} found that these transverse waves are most clearly present in Doppler shift and that their oscillation period peaks around 5 min, which are very similar to most of the type II oscillations we found here. Moreover, both these Alfv\'enic waves and our type II oscillations are persistent and no significant damping is observed. In addition, the presence of the almost vertical strips of Doppler shift features in both Figure~\ref{fig.5} and Figure~\ref{fig.6}(B) indicates that the oscillations should have a very large propagating speed. Given a phase speed of the order of 1200 km~s$^{-1}$, the wave only needs 30 s to propagate through a distance of 50$^{\prime\prime}$. This duration is comparable to the exposure time and observational cadence, which can easily explain the vertical strips. All of these observational facts suggest that some, if not all, of the type II Doppler shift oscillations might be spectroscopic signatures of the propagating Alfv\'enic waves. The periods of these oscillations, 3-6 min, suggest that the p-modes may greatly modulate the generation or propagation of these waves \citep{Tomczyk2007}. The amplitudes of these type-II Doppler shift oscillations are usually in the range of 1-2 km~s$^{-1}$, which is not that far from the velocity amplitude found by \cite{McIntosh2011} in AR using AIA data (5$\pm$5 km~s$^{-1}$). Also we have to bear in mind that the Doppler shift oscillation amplitudes observed by EIS are likely to be reduced due to the LOS integration and the coarser spatial resolution compared to AIA. For a quantification of the effect of spatial resolution on the measured Doppler shifts from Alfv\'enic waves, we refer to \cite{McIntosh2012b}. 
 
If these transverse oscillations are propagating Alfv\'enic waves, we can still calculate the internal magnetic field if the Alfv\'en speed is known. Unfortunately, the wave only needs about 30 s to propagate through a distance of 50$^{\prime\prime}$ as we mentioned above. Such a short time prevents us from deriving the phase speed from the time-distance diagram \citep{Tomczyk2008}. 

\subsubsubsection{}
As we discussed above, various observational facts seem to favor the interpretation of Alfv\'enic waves (kink or Alfv\'en waves) for our type II oscillations, although slow-mode standing waves could not be excluded for events 10 \& 21.

The WKB energy flux of these transverse waves can be estimated as following \citep[e.g.,][]{Tomczyk2007,Tomczyk2008,Ofman2008,McIntosh2011}:

\begin{equation}
\emph{$F=\rho_i(\delta v)^{2}V_{phase}$}\label{equation3},
\end{equation}

\noindent Taking the observed value of the velocity amplitude $\delta v$=1.77 km~s$^{-1}$, the calculated number density of log ({\it N}$_{i}$/cm$^{-3}$)=9.14, and the derived phase speed of 775 km~s$^{-1}$, we obtain an energy flux of 6.7$\times$10$^3$ erg cm$^{-2}$ s$^{-1}$ for event 20, which is two to three orders of magnitude lower than the energy flux required to balance the radiative and conductive losses of the active corona (2$\times$10$^6$ erg cm$^{-2}$ s$^{-1}$). But as we mentioned above, the Doppler shift oscillation amplitudes are likely to be greatly reduced due to the coarse spatial resolution and LOS integration. This is supported by recent observation showing close strands in a loop system oscillating out of phase \citep{Wang2012b}.  Recent high-resolution observations of SDO/AIA clearly reveal an oscillation amplitude of the order of 20 km~s$^{-1}$ \citep{McIntosh2011}. Such a large amplitude leads to an energy flux comparable to that required for heating the quiet corona and solar wind.  
  
\subsubsection{Intensity oscillations associated with Alfv\'enic waves}
\begin{figure*}
\centering {\includegraphics[width=\textwidth]{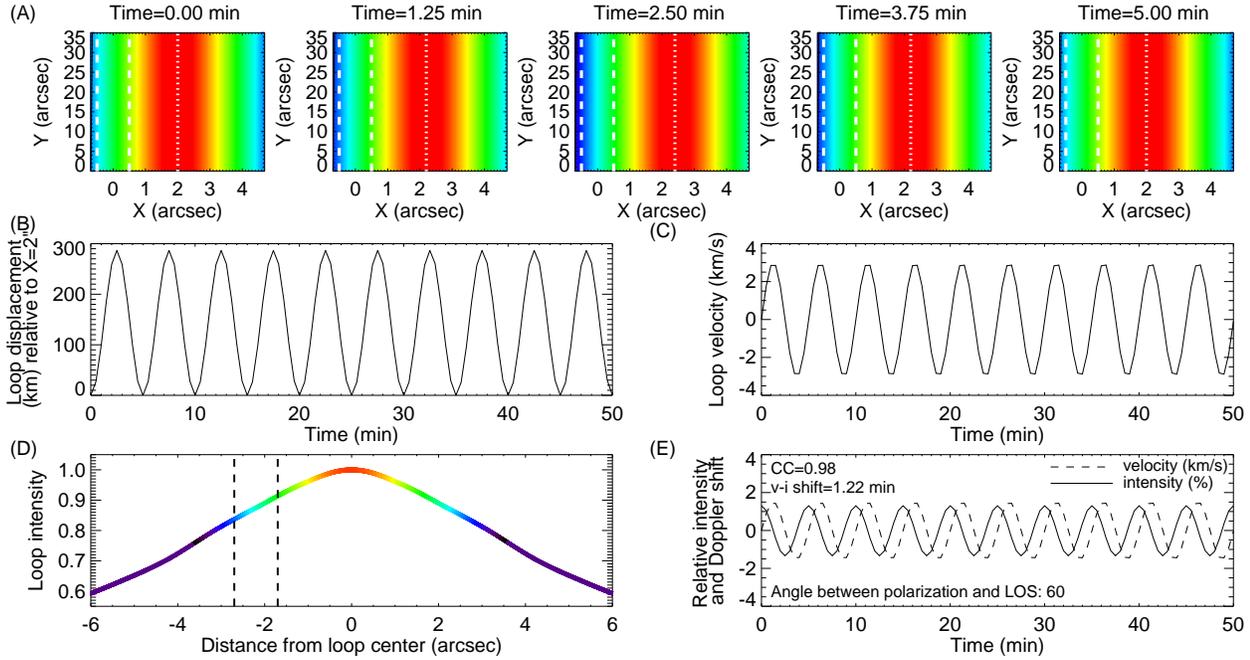}} \caption{ Intensity oscillations associated with Alfv\'enic waves. (A) Transverse displacement of a loop. X and Y are directions across and along the loop, respectively. The region between the two dashed lines (-0.5$^{\prime\prime}$$\sim$0.5$^{\prime\prime}$) indicate the slit location (or targeted pixel, 1$^{\prime\prime}$ wide). The slit is assumed to be parallel to the loop. The dotted line marks the location of the loop center. The color coding is shown in panel (D). (B) Loop displacement relative to X=2$^{\prime\prime}$. The period is 5 min. (C) Loop velocity. (D) Intensity profile across the loop. The part of the loop between the two dashed lines indicates the slit location at time=1.25 min. (E) Time series of the  intensity fluctuation and Doppler shift when the angle between the polarization direction and LOS is 60$^{\circ}$. The time lag between intensity and Doppler shift oscillations and their maximum correlation coefficient are also marked in the panel. An online movie (m1.mpg) is associated with this figure. } 
\label{fig.9}
\end{figure*}

Weak intensity oscillations could also occur in the case of Alfv\'enic waves, since periodic loop displacement could lead to the scenario that different parts (with different intensity) of a loop are sampled periodically. If the polarization of the swaying motion is such that both a LOS component (visible in Doppler shifts) and a component in the plane of the sky occur, then loops moving into and out of a spatial pixel could lead to slight intensity fluctuations that are not a signature of density changes.

Such a scenario can be easily modeled with a toy model. Figure~\ref{fig.9} shows how the intensity fluctuation, as well as a $\pi$/2 phase shift between intensity and Doppler shift can be produced. We assume a loop carrying a simple harmonic Alfv\'enic wave with an amplitude of 3 km~s$^{-1}$ and a period of 5 min (panel C). Such a wave will lead to a swaying motion which causes a periodic transverse displacement of the loop by up to $\sim$287 km (panel B). As a result of this displacement, different parts of the loop with different intensities are sampled by the instrument slit at different time (panel A). Periodic displacement leads to periodic sampling of the same part of the loop, thus producing periodic variation of the observed intensity. We derived intensity profiles across different loops from several raster scans of EIS (observed on 2007 Feb 21, 2007 May 19 and 2007 Aug 23) and present in panel (D) a typical normalized intensity profile of a loop observed by EIS. Using this intensity profile and assuming an angle of 60$^{\circ}$ between the polarization direction (X in panel A) and LOS, we obtain an intensity fluctuation of $\sim$1.5 \% (panel E) if the targeted pixel is $\sim$2$^{\prime\prime}$ away from the loop center. The Doppler shift oscillation has an amplitude of $\sim$1.5 km~s$^{-1}$ and reveals a clear $\pi$/2 phase shift with respect to the intensity oscillation. 

Intensity oscillations are usually believed to be signatures of compressible waves based on the assumption that the intensity change is a reflection of density change \citep[e.g.,][]{Kitagawa2010}. A $\pi$/2 phase shift between intensity and Doppler shift is often believed to be an indicator of standing slow waves \citep[e.g.,][]{Sakurai2002,Kitagawa2010}. However, from our Figure~\ref{fig.9} it is clear that intensity fluctuations might also be caused by sampling changing portions of a loop, with different densities, due to the periodic transverse motion. In this case, a $\pi$/2 phase shift between intensity and Doppler shift is expected if the parts of loop sampled by the instrument slit have a monotonically decreasing or increasing intensity profile. Note that the $\pi$/2 phase shift here is caused by the fact that different parts of the loop are sampled at different time, and is certainly not an intrinsic property of the transverse wave. We emphasize that for mode identification when using EIS-like spectroscopic observations the $\pi$/2 phase shift alone is not sufficient and one needs to know information such as the associated loop structure and phase speed (see section 4.2.1) for a more definite identification. 
%In disk observations of a moderate-resolution instrument like EIS, such information might sometimes 

The amplitude of intensity oscillation depends on the slope of the intensity profile in the parts of the loop sampled by the instrument slit and the angle between the polarization direction and LOS. The online movie (m1.mpg) shows how these two factors impact the amplitudes of intensity and velocity oscillations. Using a fixed angle between polarization and LOS (60$^{\circ}$), the first part of the movie shows that the intensity fluctuation becomes larger when the sampled part of the loop has a steeper intensity gradient. Note that the red line marks the central location of the sampled parts of the loop. If this sampled location is the loop center, the intensity fluctuation is small and the period of the intensity oscillation (if significant) is half of the velocity period, since the intensity slightly decreases on both sides of the loop center. The second part of the movie clearly shows that the intensity amplitude increases as the angle between polarization and LOS increases from 0$^{\circ}$ to 90$^{\circ}$. The velocity amplitude changes in the opposite sense. Except for the cases of 0$^{\circ}$ and 90$^{\circ}$, the $\pi$/2 phase shift is always there.

As we mentioned above, the intensity oscillation and the $\pi$/2 phase shift between intensity and Doppler shift variation are much more prominent in hotter lines such as Fe~{\sc{xiv}}~274.20\AA{} and Fe~{\sc{xv}}~284.16\AA{} (e.g., Figure~\ref{fig.4}). This might be explained by considering the possible different intensity profiles across the loop for different lines. If the average temperature of the surrounding coronal plasma is comparable to the formation temperature of Fe~{\sc{xii}}~195.12\AA{} or Fe~{\sc{xiii}}~202.04\AA{}, the intensity profile should be flatter for these cooler lines and thus the possible intensity fluctuation associated with the transverse oscillations will be smaller. For hotter lines, there will be a larger contrast (steeper intensity profile) between the loop and the surrounding (cooler) corona and hence the intensity fluctuation will be larger.  

The maximum correlation coefficients are always 0.98 in our idealized toy model. In real observations, the sampled parts of the loop may not always have a smooth monotonically decreasing or increasing intensity profile. This is likely to be the case as a loop is believed to consist of several sub-resolution strands. This effect will de-smooth the intensity time series and thus reduce the correlation between the intensity and Doppler shift. In addition, we assume that the oscillating part of the loop is parallel to the instrument slit, which is certainly not always the case in observations. If the loop makes a large angle with respect to the slit and there is no change of density along the loop, the observed intensity change will be reduced and the $\pi$/2 phase shift might be less clear. In fact we see a  more obvious $\pi$/2 phase shift in events 10, 17, 19, 20 and 21. In all of these events the top part of the oscillating loop is more or less parallel to the EIS slit. In other type-II oscillation events the oscillating loop makes a large angle with respect to the slit and we generally do not see a clear $\pi$/2 phase shift. Another major factor that would cause a reduced correlation, a variable phase shift and intensity oscillation amplitude is the possible change of the sampled loop location throughout the timeseries. This could happen due to either natural evolution of the loop or slight wobble/jitter in the EIS pointing. Finally, the instrument noise will certainly reduce the correlation coefficient and cause a variable phase shift with time. 

\section{Conclusion}

We have performed a statistical study of Doppler shift oscillations using data taken by HINODE/EIS. We have found mainly two types of oscillations: one type (type I) is mainly found at loop footpoint regions and the other (type II) is typically associated with the upper part of loops. The type I oscillations generally show coherent behavior of all line parameters (line intensity, Doppler shift, line width and profile asymmetry), apparent blue shift and blueward asymmetry throughout almost the entire observational duration. These observational facts seem to be consistent with the scenario of episodic high-speed upflows (fine-scale recurrent jets). The type II oscillations, with a period of 3-6 minutes, are most clearly seen in the Doppler shift and often show no significant variation of the intensity and line width. The line profiles do not show any obvious asymmetry. These are probably spectroscopic signatures of kink/Alfv\'en waves. In addition, we have presented a toy model, demonstrating that such transverse waves could also produce an observed intensity change and a $\pi$/2 phase shift between the intensity and Doppler shift oscillations if the associated loops have monotonic density variation. Thus, the value of the phase shift alone is not a reliable diagnostic of the wave mode in spectroscopic observations made by instruments such as HINODE/EIS.

\begin{acknowledgements}
EIS is an instrument onboard {\it Hinode}, a Japanese mission developed and launched by ISAS/JAXA, with NAOJ as domestic partner and NASA
and STFC (UK) as international partners. It is operated by these agencies in cooperation with ESA and NSC (Norway). S. W. McIntosh is supported by NASA (NNX08AL22G, NNX08BA99G) and NSF (ATM-0541567, ATM-0925177). T. Wang and L. Ofman acknowledge supports by NASA grants NNX10AN10G and NNX12AB34G. H. Tian is supported by the
ASP Postdoctoral Fellowship Program of National Center for Atmospheric Research, which is sponsored by the National Science Foundation. H. Tian thanks I. De Moortel, G. R. Gupta and L. Teriaca for helpful discussions.

\end{acknowledgements}

\end{document}